\documentclass[
aps, 
pra, 
onecolumn, 
superscriptaddress, 
showkeys, 
]{revtex4-2}

\usepackage[english]{babel}
\renewcommand{\selectlanguage}[1]{} 

\usepackage{xcolor}
\usepackage[hidelinks]{hyperref}
\usepackage{physics}
\usepackage{graphicx}

\usepackage[capitalise,nameinlink]{cleveref} 

\usepackage{caption}
\usepackage{subcaption} 

\newtheorem{definition}{Definition}

\makeatletter
\let\cat@comma@active\@empty
\makeatother

\begin{document}

\title{Quantum simulation of wave optics in weakly inhomogeneous media using block-encoding}

\author{Siavash Davani}
\email{siavash.davani@uni-jena.de}
\affiliation{Institute of Applied Physics, Abbe Center of Photonics, Friedrich Schiller University Jena, 07745 Jena, Germany}
\affiliation{Max Planck School of Photonics, 07745 Jena, Germany}

\author{Martin G{\"a}rttner}
\email{martin.gaerttner@uni-jena.de}
\affiliation{Institute of Condensed Matter Theory and Optics, Friedrich Schiller University Jena, 07743 Jena, Germany}

\author{Falk Eilenberger}
\email{falk.eilenberger@uni-jena.de}
\affiliation{Fraunhofer Institute for Applied Optics and Precision Engineering IOF, 07745 Jena, Germany}
\affiliation{Institute of Applied Physics, Abbe Center of Photonics, Friedrich Schiller University Jena, 07745 Jena, Germany}
\affiliation{Max Planck School of Photonics, 07745 Jena, Germany}

\date{May 28, 2026; Revised: September 22, 2026}

\begin{abstract}
We propose a quantum algorithm that simulates the propagation of a light field through a weakly inhomogeneous medium. In the paraxial approximation, the wave equation in an inhomogeneous material takes the form of the Schrödinger equation with a time-dependent Hamiltonian. This reduction is used to simulate wave optical dynamics on a quantum computer. Beam propagator operators for a short propagation distance are constructed using an efficient and flexible block-encoding that enables the simulation of various optical setups. The algorithm is showcased by simulating the propagation of a one-dimensional Gaussian beam through a lens of finite thickness, and the resulting spherical aberration is demonstrated.
\end{abstract}

\maketitle

\section{Introduction}\label{sec-introduction}

In recent years, quantum computers have received significant attention because of their potential to advance various fields and technologies including simulation \cite{daley_practical_2022, altman_quantum_2021}, cryptography and communications \cite{pirandola_advances_2020}, machine learning \cite{cerezo_challenges_2022}, and many more \cite{dalzell_quantum_2025, santagati_drug_2024, herman_quantum_2023, di_meglio_quantum_2024, mazzola_quantum_2024}. In particular, the simulation of physical systems was among the very first proposed applications of quantum computers \cite{feynman_simulating_1982} and has remained central to quantum computing research ever since \cite{lloyd_universal_1996, georgescu_quantum_2014}.

Quantum algorithms have been introduced for solving linear systems \cite{harrow_quantum_2009} and differential equations \cite{berry_high-order_2014}. In some cases, these algorithms can exponentially improve the performance of simulations compared to their classical counterparts \cite{dalzell_quantum_2025}, once a fault-tolerant quantum computer is available \cite{webster_universal_2022, toshio_practical_2025, battistel_real-time_2023, campbell_roads_2017}.
On the algorithmic side, much work has been done on fundamental primitives such as linear combination of unitaries (LCU) \cite{childs_hamiltonian_2012}, qubitization \cite{low_hamiltonian_2019}, quantum signal processing (QSP), and quantum singular value transformation (QSVT) \cite{gilyen_quantum_2019}. These efforts have converged to a unified formalism \cite{martyn_grand_2021}, gathering many of the known quantum algorithms under one common framework using the notion of block-encoding \cite{low_hamiltonian_2019}. Quantum signal processing using block-encoding has been shown to achieve optimal computational complexity in the time-independent Hamiltonian simulation problem \cite{low_hamiltonian_2019, gilyen_quantum_2019}.

Nevertheless, a robust quantum algorithm for simulating wave optics has not been proposed so far. The diverse applications of wave propagation algorithms range from designing optical waveguides, lasers, and integrated circuits, to biomedical optics and imaging \cite{schmidt_semi-analytic_2022,heintzmann_scalable_2023, wende_fast_2024, liu_wave_2024, sung_realistic_2025}. In modern optical engineering, there is an increasing need for more accurate simulations of beam propagation in complex structures where the precise control of light is essential. These simulations require significant memory and processing power; improving their performance is therefore highly desired \cite{lu_systolic_2025, wu_accelerated_2026, shams_optimization_2011}. Quantum computers are potential candidates to reduce these computational costs. Quantum registers store a field discretized at $N$ spatial points in only $\log_2 N$ qubits. This efficiency in storage can potentially lead to exponential improvements in the performance as well, since quantum algorithms operate on a logarithmically smaller number of memory units compared to classical algorithms. Finding such efficient quantum algorithms for domain-specific applications is a key objective of quantum simulation research.

Previous works presented quantum algorithms for calculating the electromagnetic scattering cross section \cite{clader_preconditioned_2013}, performing quantum beam propagation in homogeneous media \cite{cholsuk_efficient_2024}, and simulating quantum optical systems \cite{kottmann_quantum_2021}.
In this work, we simulate wave optics on quantum computers by solving the Helmholtz equation for a scalar field in a medium with a slowly varying refractive index. The simulator is implemented using an efficient and flexible block-encoding structure that can be tuned to simulate the effects of various optical elements on an incident light beam. The implementation uses an ancillary quantum register to construct block-encoded phase shifters acting on a target register. Using a split-step technique \cite{burzler_split-step_1996}, these phase shifters are used along with quantum Fourier transform (QFT) subroutines \cite{nielsen_quantum_2012} to simulate wave propagation in the paraxial approximation. The ancillary register only needs $\log_2 N$ qubits, and the gate complexity of the simulator up to an error $\epsilon$ defined using the spectral norm is quadratic in the simulation time $\mathcal{O}(T^2/\epsilon)$. Therefore, this algorithm establishes a significant improvement in computational capacity for high-precision applications.

\cref{sec-preliminary} discusses the reduction of the Helmholtz equation to a time-dependent Hamiltonian simulation problem. In \cref{sec-results}, the block-encoding of phase propagators is introduced and its relation to Hamiltonian simulation is discussed. Finally, \cref{sec-sim} presents the simulation of a Gaussian beam incident on a plano-convex lens using the phase propagators.

\section{Preliminaries}\label{sec-preliminary}

\subsection{Paraxial wave optics in weakly inhomogeneous media as a Hamiltonian simulation problem}

The scalar wave equation for a field $u(\vb r,\omega)$ propagating in a weakly inhomogeneous medium is
\begin{equation}\label{eq-scalar-field-hom}
    \nabla^2 u(\vb r, \omega) + k^2(\vb r, \omega) u(\vb r, \omega) = 0,
\end{equation}
known as the Helmholtz equation \cite{saleh_fundamentals_1991}, where
\begin{equation}\label{eq-wavenumber-with-r}
    k^2(\vb r, \omega)=\frac{\omega^2}{c^2} \varepsilon_r(\vb r, \omega)
\end{equation}
is the squared wavenumber. Here, $\omega$, $\varepsilon_r$, $c$, and $\nabla^2$ are, respectively, the angular frequency of the monochromatic wave, the relative permittivity of the medium, the speed of light in vacuum, and the Laplacian operator. \cref{eq-scalar-field-hom} describes the propagation of a wave in an isotropic, dispersive, and inhomogeneous material as long as $\varepsilon_r(\vb r, \omega)$ varies slowly as a function of the spatial coordinate $\vb r$, that is
\begin{equation}
    \abs{\frac{\nabla\varepsilon_r(\vb r, \omega)}{\varepsilon_r(\vb r, \omega)}} \ll \frac{1}{\lambda},
\end{equation}
where $\lambda$ is the wavelength.
In our discussion, we consider the case of a monochromatic wave and leave out the $\omega$ dependence of $\varepsilon_r$ from now on.
We also assume that the radiation is localized in the momentum space and apply the paraxial approximation. Without loss of generality, we choose the $z$ coordinate axis to align with the optical axis. After these assumptions, the solution to \cref{eq-scalar-field-hom} takes the form
\begin{equation}
    u(\vb r) = v(\vb r) e^{i\tilde{k}z},
\end{equation}
where $\tilde{k}=\expval{k(\vb r)}$ is the wavenumber averaged over the volume of the simulation space. Inserting this ansatz into \cref{eq-scalar-field-hom} leads to the following differential equation for $v(\vb r)$
\begin{equation}\label{eq-wave-equation-approximated}
    i \pdv{}{z}v(\vb r) + \frac{1}{2\tilde{k}}\nabla_\perp^2 v(\vb r) + \frac{k^2(\vb r)-\tilde{k}^2}{2\tilde{k}} v(\vb r) = 0,
\end{equation}
where $\nabla_\perp^2 = \pdv[2]{}{x}+\pdv[2]{}{y}$ is the transverse Laplacian operator. Since we chose the $z$ axis to match the direction of propagation, we can separate the transverse coordinates $\vb r_\perp=x \vb e_x+y \vb e_y$ and the propagation coordinate $z=ct$ to rewrite the differential equation as
\begin{equation}
    i\frac{1}{c}\pdv{}{t}v(t; \vb r_\perp) + \frac{1}{2\tilde{k}}\nabla_\perp^2 v(t; \vb r_\perp) + \frac{k^2(t; \vb r_\perp)-\tilde{k}^2}{2\tilde{k}} v(t; \vb r_\perp) = 0.
\end{equation}
This equation is a first-order differential equation in time and resembles the form of the Schrödinger equation
\begin{equation}\label{eq-schorodinger-for-propagation}
    i\hbar\pdv{}{t}\ket{v(t)} = \hat H(t) \ket{v(t)},
\end{equation}
where
\begin{equation}
    \braket{\vb r_\perp}{v(t)} = v(t; \vb r_\perp),
\end{equation}
and the Hamiltonian of the system is
\begin{align}
    \hat H(t) &= \frac{c}{2\hbar\tilde{k}}(\hat p_x^2+\hat p_y^2) - \frac{\hbar c}{2\tilde{k}} (k^2(t; \hat{x}, \hat{y})-\tilde{k}^2) \\
    &= K(\hat p_x,\hat p_y) + V(t; \hat{x}, \hat{y}) \label{eq-ham-t-v-split},
\end{align}
where $K$ and $V$ are the kinetic and potential energy terms. This shows that the problem of beam propagation in weakly inhomogeneous media reduces to a time-dependent Hamiltonian simulation problem and hence can be solved using a quantum computer. Note that the potential energy term in the Hamiltonian is time-dependent (or equivalently $z$-dependent in this case) because the propagating beam interacts with a varying refractive index during the propagation.

A simulation algorithm starts with a field distribution in the transverse plane $\ket{v(0)}$ at the initial time $t=0$ and evolves the field after propagating for time $T$, or distance $z=cT$, in the direction of propagation to compute $\ket{v(T)}$. The time dependence of the Hamiltonian and the non-commuting kinetic and potential energy terms in \cref{eq-ham-t-v-split} make this computation non-trivial. We use a split-step method \cite{burzler_split-step_1996} to perform the computation on a quantum computer by dividing the evolution time into a number of small time steps and executing the corresponding unitary steps sequentially.

Let us partition the total evolution time from $t=0$ to $t=T$ into $N_t$ equal intervals $\Delta t = T/N_t$ and define the time grid $t_j=j\Delta t$. Assuming $\Delta t$ is small and using the first-order Trotter-Suzuki formula \cite{trotter_product_1959}, we can approximate the total time evolution operator (setting $\hbar=1$) as a sequence of evolution operators for time $\Delta t$. For each time step we use a midpoint approximation by choosing $t_{j+1/2}$ to evaluate the Hamiltonian in the time window $[t_j, t_{j+1}]$ which leads to
\begin{equation}\label{eq-time-ordered-evolution-approx}
    \hat U(T,0) = \prod_{j=0}^{N_t-1} e^{-i\hat K\Delta t} e^{-i\hat V(t_{j+1/2})\Delta t} + \mathcal{O}\Big( \frac{T^2}{N_t} \Big).
\end{equation}
This means, to execute the time evolution, we need to sequentially apply unitary propagation steps of the form $e^{-i\hat K\Delta t}$ and $e^{-i\hat V(t)\Delta t}$ to the state $\ket{v}$. The kinetic energy part $\hat K$ is not a function of time and is a simple quadratic function of the transverse frequency. Therefore, $e^{-i\hat K\Delta t}$ can be efficiently implemented using standard diagonal unitary synthesis and QFT subroutines \cite{cholsuk_efficient_2024}. However, the potential energy part $V(t; \hat x, \hat y)$ is in general changing with propagation time and can be an arbitrary function of transverse coordinates.
In \cref{sec-results}, we construct block-encoded operators that implement such unitary potential energy propagators and use them to realize $\hat U(T,0)$.
The block-encoding is efficient for implementing a phase operator $\exp(if(\hat x))$ under two conditions: (i) the norm of the phase $\norm{f(x)}_1$ should be small; and (ii) there should be an efficient state preparation unitary for the state $\phi(x)$ such that $f(x) \propto \abs{\phi(x)}^2$.
Condition (i) is naturally fulfilled in the wave propagation problem discussed here, because in order for the split-step approximation in \cref{eq-time-ordered-evolution-approx} to be valid, the time step $\Delta t$ needs to be small, leading to small local phases $f(x)=-V(t;x)\Delta t$ at each time step. Condition (ii) remains as a requirement for efficiency, meaning that the proposed simulator will be efficient if the geometry of the simulation problem leads to refractive index distribution profiles for which we can find efficient state preparation unitaries.
Although state preparation is not efficient in general, there are large classes of states relevant to optics (e.g., smooth functions with bounded derivatives) that have efficient initializers \cite{ran_encoding_2020,iaconis_quantum_2024,rosenkranz_quantum_2025,moosa_linear-depth_2023}. Therefore, the block-encoding construction is suitable for implementing the evolution operator in \cref{eq-time-ordered-evolution-approx}. These conditions are discussed further in the following sections and the detailed analysis of the error scaling is included in Appendix~\ref{sec:appendix_simulation_error_analysis}.

It is also worth mentioning that we have used the first-order Trotter formula in this discussion because of its simplicity. However, it is possible to use more advanced methods such as higher-order Trotterization \cite{yang_improved_2022} or Dyson series \cite{kieferova_simulating_2019} to achieve a more favorable scaling of the time discretization error with $\Delta t$. Our main objective in this work, however, is to implement evolution operators for a small step size using block-encoding, which then can also be used in other time discretization schemes.

\subsection{Block-encoding}

Before presenting the main results, we briefly review the notion of block-encoding. Block-encoding is an intuitive and useful way to model non-unitary operations on quantum computers. The formal definition of block-encoding is given in the following. The presentation follows \cite{gilyen_quantum_2019}.

\begin{definition}[Block-encoding]
    A unitary operator $U_A$ is called an $(\alpha, a, \epsilon)$-block-encoding of the (not necessarily unitary) operator $A$ if
    \begin{equation}
    \norm{
        A - \alpha\Big(\bra{0}^{\otimes a}\otimes I\Big)U_A\Big(\ket{0}^{\otimes a}\otimes I\Big)
    } \le \epsilon\,.
    \end{equation}
\end{definition}
As an intuitive example of the above definition, an exact block-encoding of $A$ ($\epsilon=0$) takes the form
\begin{equation}
U_A=\begin{pmatrix}
  A/\alpha & *\\
  * & *
\end{pmatrix},
\end{equation}
meaning that $A$ occupies the top left block of $U_A$, and $*$ indicates that the exact values of the other blocks of $U_A$ are not important as long as $U_A$ is a unitary operator. If $A$ were unitary and we knew how to construct it on a quantum computer, we could directly apply it to a register $\ket{\psi}$ to achieve the result
\begin{equation}
    A\ket{\psi}.
\end{equation}
But since $A$ can be non-unitary, $A\ket{\psi}$ is not necessarily normalized, and thus we can only hope for probabilistic access to it. The block-encoding $U_A$ provides such probabilistic access. In this example, an $a$-qubit ancillary register is added to the primary register to form $\ket{0}^{\otimes a}\ket{\psi}$ and the block-encoding $U_A$ acts on the combined register
\begin{equation}
    U_A\ket{0}^{\otimes a}\ket{\psi} = \ket{0}^{\otimes a} \frac{A\ket{\psi}}{\alpha} + \ket{\perp},
\end{equation}
where $(\bra{0}^{\otimes a}\otimes I)\ket{\perp}=0$. If we post-select on the $a$-qubit ancillary register being in the zero state, the primary register is projected onto a state proportional to $A\ket{\psi}$, meaning that the ancillary register flags the success case. The probability of success of the post-selection is
\begin{equation}
P_{\mathrm{success}} = \frac{1}{\abs{\alpha}^2} \bra{\psi} A^\dagger A \ket{\psi}.
\end{equation}
Note that any unitary operator is trivially a $(1,0,0)$-block-encoding of itself; hence, using the block-encoding framework, we can extend the analysis of quantum algorithms beyond unitary operations by allowing for ancillary registers, probabilistic success, and flagging of the success case.

\section{Results}\label{sec-results}

\begin{figure}[ht]
    \centering
    \includegraphics[width = 0.4\textwidth]{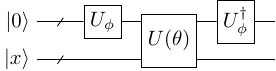}
    \caption{Schematic circuit representation of the block-encoding unit for the phase protocol.}
    \label{fig-block-encoding}
\end{figure}

In order to implement potential energy propagators in the time evolution operator in \cref{eq-time-ordered-evolution-approx}, we first construct a block-encoding unit for a generic phase operator
\begin{equation}\label{eq-unit-block-encoding}
    e^{i\sum\limits_x\Delta\abs{\phi(x)}^2 \ketbra{x}{x}}
\end{equation}
for a small coefficient $\Delta$ and a normalized wavefunction $\phi(x)$. Then we discuss under what conditions this block-encoding unit can be repeated to synthesize the phase propagator
\begin{equation}
    e^{i\sum\limits_x\alpha\abs{\phi(x)}^2 \ketbra{x}{x}}
\end{equation}
for larger coefficients $\alpha$ with high accuracy. Because the unitary propagators $e^{-iV(t;\hat x)\Delta t}$ from \cref{eq-time-ordered-evolution-approx} are diagonal phase operators in the position basis and the phase $V(t; x)\Delta t$ is small at each time step, we expect this block-encoding structure to efficiently implement the propagator using only a small number of iterations by choosing appropriate $\Delta$ and $\phi(x)$ parameters. Hence, together with the kinetic propagators, the block-encoding unit realizes a single time step of the unitary evolution for a small step size, and this structure is then used in iterations to simulate arbitrarily long dynamics. The circuit representation of the block-encoding unit is presented in \cref{fig-block-encoding}.
Here, $U_\phi$ is an amplitude oracle unitary that prepares the state $\ket{\phi}$ in the $n_q$-qubit ancillary register
\begin{equation}
    U_\phi\ket{0}=\ket{\phi} = \sum_{x=0}^{2^{n_q}-1} \phi(x) \ket x,
\end{equation}
and $U(\theta)$ is defined as a conditional phase operator
\begin{equation}
    U(\theta) \ket{y}\ket{x} = \begin{cases}
    e^{i\theta} \ket{y}\ket{x}, & \text{for } x=y \\
    \ket{y}\ket{x}, & \text{for } x \ne y
    \end{cases}.
\end{equation}
The block-encoding uses the ancillary state $\ket\phi$ to apply the $e^{i\Delta\abs{\phi(x)}^2}$ phase to the $n_q$-qubit primary register $\ket{x}$.

As we will show in the following, the $U(\theta)$ operator is very efficient, requiring only $\mathcal{O}(n_q)$ gates for its implementation. Therefore, the efficiency of the block-encoding is determined by the gate complexity of the $U_\phi$ operator. The main advantage of using this block-encoding structure is that it reduces the problem of diagonal phase operator construction to finding $U_\phi$ initializers for the corresponding phase distribution.
Hence, the efficiency of our algorithm hinges upon efficient initializer circuits $U_\phi$.
However, the strength of this approach is that any phase profile, for which the corresponding state amplitudes can be efficiently prepared, is directly mapped to a unitary phase operator and can be treated efficiently by our algorithm to simulate wave propagation in a geometry with a given refractive index structure.

The construction of state preparation unitaries for specific states is a topic of ongoing research, and a full characterization of the efficiently preparable classes of functions remains elusive.
In optics and many other scientific applications, the phase distributions usually belong to the class of smooth functions with bounded derivatives, and one can use techniques based on matrix product states \cite{ran_encoding_2020, iaconis_quantum_2024} to prepare the states efficiently. Specific state initializers for particularly common states such as Gaussian distributions have also been reported in the literature \cite{rosenkranz_quantum_2025}. Periodic distributions such as sine- or binary-type gratings can also be implemented by first preparing the Fourier spectrum of the function followed by an inverse QFT \cite{moosa_linear-depth_2023}.  For an overview of the state preparation problem and the representation of data on a quantum computer, see \cite{schuld_representing_2021, dalzell_quantum_2025}. These examples already show that many interesting real-world applications can be implemented using our algorithm, and the approach directly benefits from advances in the state preparation problem.

\begin{figure}[ht]
    \centering
    \includegraphics[width = 0.5\textwidth]{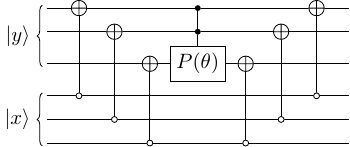}
    \caption{Implementation of the conditional phase operator $U(\theta)$ for $3$-qubit input registers, using zero-controlled CNOT gates and a doubly-controlled phase gate.}
    \label{fig-partial_phase_operator_example}
\end{figure}

The role of the $U(\theta)$ operator is to establish the necessary entanglement between the registers to achieve the desired phase transformation. This operator is implemented using $\mathcal{O}(n_q)$ elementary gates. The general implementation of the operator for arbitrary $n_q$ is discussed in Appendix~\ref{sec:appendix_implementation}. However, as an example, the implementation for $n_q=3$ is presented in \cref{fig-partial_phase_operator_example}, where
$P(\theta)=\left[\begin{smallmatrix} 1 & 0 \\ 0 & e^{i\theta} \end{smallmatrix}\right]$ is the elementary phase operator.

The efficient $\mathcal{O}(n_q)$ gate complexity means that, in most cases, the $U(\theta)$ operator will not be the limiting factor in the performance of the protocol, considering the usually higher gate complexity of the state preparation unitary $U_\phi$. Also, when the protocol is used together with the QFT, e.g., in most simulation problems, the performance will be limited by the higher $\mathcal{O}(n_q^2)$ gate complexity of the QFT.

Direct calculation (Appendix~\ref{sec:appendix_outcome_calculations}) of the output of the circuit in \cref{fig-block-encoding} shows that the circuit is a $(1,n_q,0)$-block-encoding of the operator
\begin{equation}
    A(\theta) = I + 2ie^{i\frac{\theta}{2}}\sin\frac{\theta}{2} \sum\limits_x\abs{\phi(x)}^2 \ketbra{x}{x}
\end{equation}
such that the following transformation is performed by the circuit
\begin{equation}
    U_A\ket{0}^{\otimes n_q}\ket{\psi} = \ket{0}^{\otimes n_q} A(\theta)\ket{\psi} + \ket{\perp}.
\end{equation}
The success probability of the operation (namely the probability of the ancillary register being $\ket{0}$) is
\begin{align}
    P_{\mathrm{success}}
    &=\bra{\psi}A^\dagger(\theta)A(\theta)\ket{\psi} \nonumber \\
    &\ge 1-\sin^2\frac{\theta}{2}.
\end{align}
If we take the phase to be small, $\theta=\Delta \ll 1$, then the operator becomes 
\begin{align}
    A(\Delta)
    &= I + i\Delta\sum\limits_x\abs{\phi(x)}^2 \ketbra{x}{x} \nonumber + \mathcal{O}(\Delta^2) \\
    &= e^{i\Delta\sum\limits_x\abs{\phi(x)}^2 \ketbra{x}{x}} + \mathcal{O}(\Delta^2) \label{eq-block-encoding-error}
\end{align}
in the first-order approximation in $\Delta$, where the success probability of the operation is
\begin{equation}
    P_{\mathrm{success}} = 1 - \mathcal{O}(\Delta^2).
\end{equation}
Starting with the primary register in the state $\ket{\psi}=\sum_x\psi(x)\ket{x}$ and applying the block-encoding to it, we effectively imprint a phase onto the state as
\begin{equation}\label{eq-phase_transformation_step3}
\ket{\psi}=\sum_x\psi(x)\ket{x}
\,\longrightarrow\,
\ket{\psi'}\approx\sum_x\psi(x)\,e^{i\Delta\abs{\phi(x)}^2}\ket{x}
\end{equation}
with arbitrarily high probability. The circuit in \cref{fig-block-encoding} is thus a $(1,n_q,\mathcal{O}(\Delta^2))$-block-encoding of the operator
\begin{equation}\label{eq-approx-unitary}
    e^{i\Delta\sum\limits_x\abs{\phi(x)}^2 \ketbra{x}{x}}.
\end{equation}
Hence, if we post-select the success case and repeatedly apply the circuit $m$ times, we implement the operator
\begin{equation}\label{eq-repeated-unitary}
    e^{i\sum\limits_x\alpha\abs{\phi(x)}^2 \ketbra{x}{x}},
\end{equation}
where $\alpha=m\Delta$, and the error of this implementation is $\epsilon=\mathcal{O}(m\Delta^2)=\mathcal{O}(\alpha^2/m)$. Therefore, the repeated application of the block-encoding unit is efficient when the phase coefficient $\alpha$ is not too large. Given a phase coefficient $\alpha$, the number of iterations needed to achieve an error less than $\epsilon$ is $m=\mathcal{O}(\alpha^2/\epsilon)$.

This allows for the implementation of the time evolution operator in \cref{eq-time-ordered-evolution-approx} as
\begin{align}
    \hat U(T,0)
    &= \prod_{j=0}^{N_t-1} e^{-i\hat K\Delta t} e^{-i\hat V(t_{j+1/2})\Delta t} + \epsilon_{\mathrm{tot}},
\end{align}
where $\epsilon_{\mathrm{tot}}=\mathcal{O}( T^2 / N_t )$ is the total simulation error, and the $e^{-i\hat V(t)\Delta t}$ operators are implemented using the block-encoding construction by choosing suitable ancillary states $\ket\phi$ and phase coefficients $\alpha$ corresponding to the desired phase profiles $V(t; x)\Delta t$ at each time step.
The requirement of the protocol on the $\Delta$ parameter to be small also fits well with the assumption of small $\Delta t$ in the Trotter scheme used to simplify the total time evolution unitary, which leads to no more than $m=\mathcal{O}(1)$ iterations of the block-encoding unit to realize the potential propagator at each time step.
See Appendix~\ref{sec:appendix_simulation_error_analysis} for a detailed analysis.
This means, in order to perform a time evolution for total time $T$ with a maximum tolerated error $\epsilon_{\mathrm{tot}}$, we need $\mathcal{O}\!\left(\frac{T^2}{\epsilon_{\mathrm{tot}}}\right)$ iterations.
Also note that the block-encoding in \cref{eq-block-encoding-error} directly implements the first-order Taylor expansion of the unitary
\begin{equation}
    e^{i\sum\limits_x \Delta\abs{\phi(x)}^2 \ketbra{x}{x}}
\end{equation}
and therefore does not require using advanced Hamiltonian simulation techniques such as quantum signal processing \cite{gilyen_quantum_2019, gilyen_optimizing_2019}.

\section{Simulations}\label{sec-sim}

\begin{figure}[ht]
    \centering
    \includegraphics[width=0.33\textwidth]{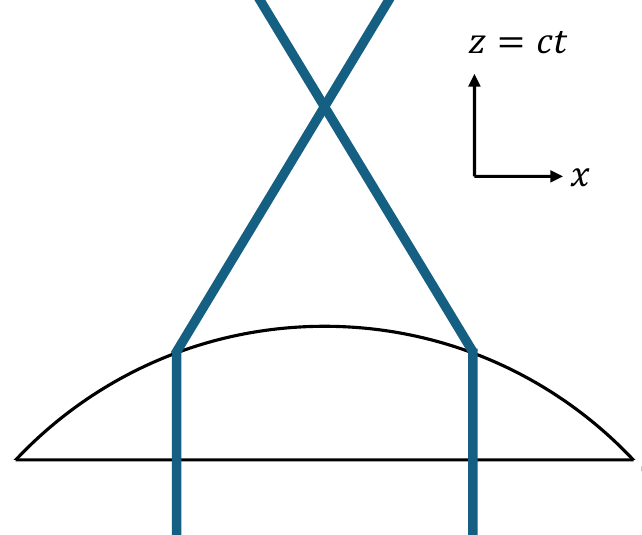}
    \caption{Schematic visualization of the simulation setup. The beam propagates in the $z$ direction.}
    \label{fig-simulation-schematic}
\end{figure}

To showcase the algorithm, we simulated the propagation of a one-dimensional Gaussian beam in a two-dimensional setup passing through a spherical (plano-convex) lens \cite{davani_2026_19333736_code, davani_2026_19333783_dataset}. We assume the paraxial approximation and align the $z$ axis with the propagation direction. The lens is expected to cause the beam to focus at a certain propagation distance after the lens. The lens has a finite non-zero thickness, and due to the spherical shape of its surface, spherical aberration is expected to appear. The lens is also assumed to be lossless; therefore, it only imprints phases on the incident beam. The schematic plot of the setup is shown in \cref{fig-simulation-schematic}.

\begin{figure}[ht]
    \centering
    \hfill
    \begin{subfigure}[t]{0.28\textwidth}
        \centering
        \includegraphics[width=\textwidth]{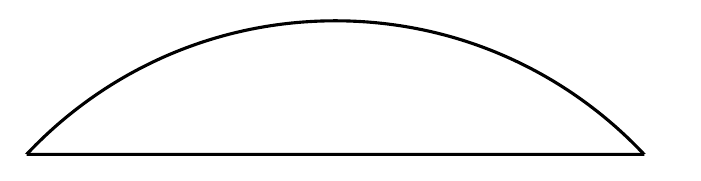}
        \caption{Original lens}
    \end{subfigure}
    \hfill
    \begin{subfigure}[t]{0.28\textwidth}
        \centering
        \includegraphics[width=\textwidth]{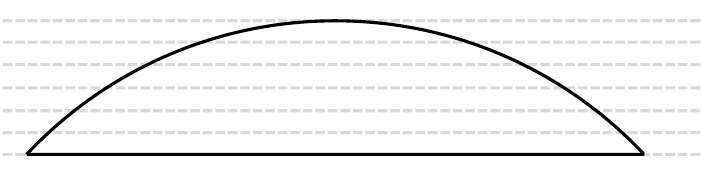}
        \caption{Slicing}
    \end{subfigure}
    \hfill
    \begin{subfigure}[t]{0.28\textwidth}
        \centering
        \includegraphics[width=\textwidth]{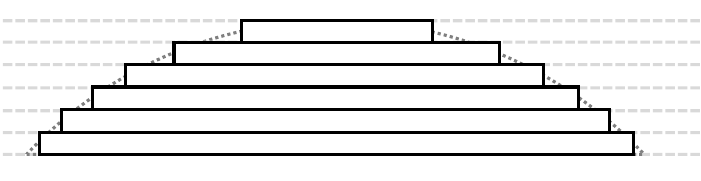}
        \caption{Rectangular approximation}
    \end{subfigure}
    \hfill
    \caption{Slicing the lens and approximating it by a stack of rectangular layers with a fixed thickness in the propagation direction.}
    \label{fig-lens-slicing}
\end{figure}

Following the split-step procedure described in \cref{eq-time-ordered-evolution-approx}, in order to perform the simulation using the protocol, we cut the lens perpendicular to the propagation axis and approximate it as a stack of thin rectangular layers of material. This procedure is shown in \cref{fig-lens-slicing}. Note that the propagation distance $z$ and the elapsed time are interchangeable according to $z=ct$.

\begin{figure}[ht]
    \centering
    \hfill
    \begin{subfigure}[t]{0.48\textwidth}
        \centering
        \includegraphics[width=\textwidth]{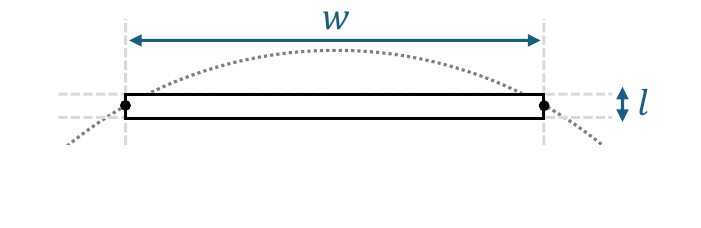}
        \caption{}
    \end{subfigure}
    \hfill
    \begin{subfigure}[t]{0.48\textwidth}
        \centering
        \includegraphics[width=\textwidth]{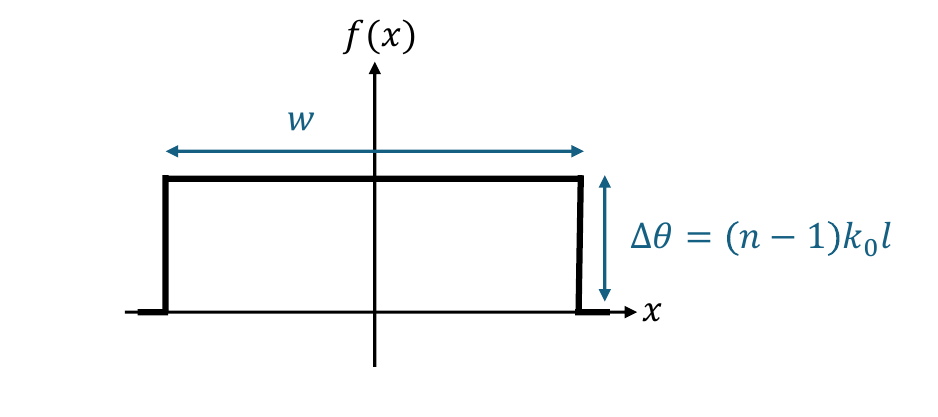}
        \caption{}
    \end{subfigure}
    \hfill
    \caption{The phase difference that each layer of the material imprints on the incident beam, depending on whether the corresponding part of the wave in the transverse axis passes through the material or propagates through the vacuum outside the material. (a) The geometry of a slice of the lens. (b) The corresponding relative phase $e^{if(x)}$ that each slice of the lens imprints on the incident beam in the real domain.}
    \label{fig-lens-phase-shifts}
\end{figure}

This allows us to propagate the beam through the lens step-by-step by sequentially applying a rectangular phase profile to the field (in the real domain) and propagating it for the thickness of each layer (quadratic phase profile in the Fourier domain). For the phase corresponding to each layer of material in the real domain, the ancillary state $\ket\phi$ is chosen as a rectangular state whose width is calculated using the curvature of the lens surface, and the phase coefficient $\alpha$ is chosen such that the part of the field passing through the material picks up a phase equal to $\Delta\theta=(n-1)k_0l$, where $n$, $k_0$, and $l$ are, respectively, the refractive index of the material, the vacuum wavenumber, and the thickness of each slice (\cref{fig-lens-phase-shifts}).
This particular approximation is not the only possible choice and is used here for its simplicity. We note that the ability of the phase protocol to tune the applied phase profiles provides great flexibility to design and explore various possibilities for such simulations. This flexibility is showcased in Appendix~\ref{sec:appendix_simulation}, where several other lens structures are simulated.

Moreover, the kinetic energy part of the Hamiltonian that simulates the propagation of the field along the $z$ axis is implemented in three steps \cite{cholsuk_efficient_2024}: (1) applying a QFT to the state $\ket{v}$, (2) applying the paraxial-approximated transfer function (a quadratic phase) to the Fourier transform of the field as
\begin{equation}
    \mathcal{H}_f(k_\perp)=e^{-i\frac{k_\perp^2}{2\tilde{k}}z},
\end{equation}
where $k_\perp$ is the transverse spatial frequency and $z$ is the propagation distance, and finally (3) applying an inverse QFT to retrieve the propagated field in the real domain.

\begin{figure}[t]
    \centering
    \includegraphics[width=0.65\textwidth]{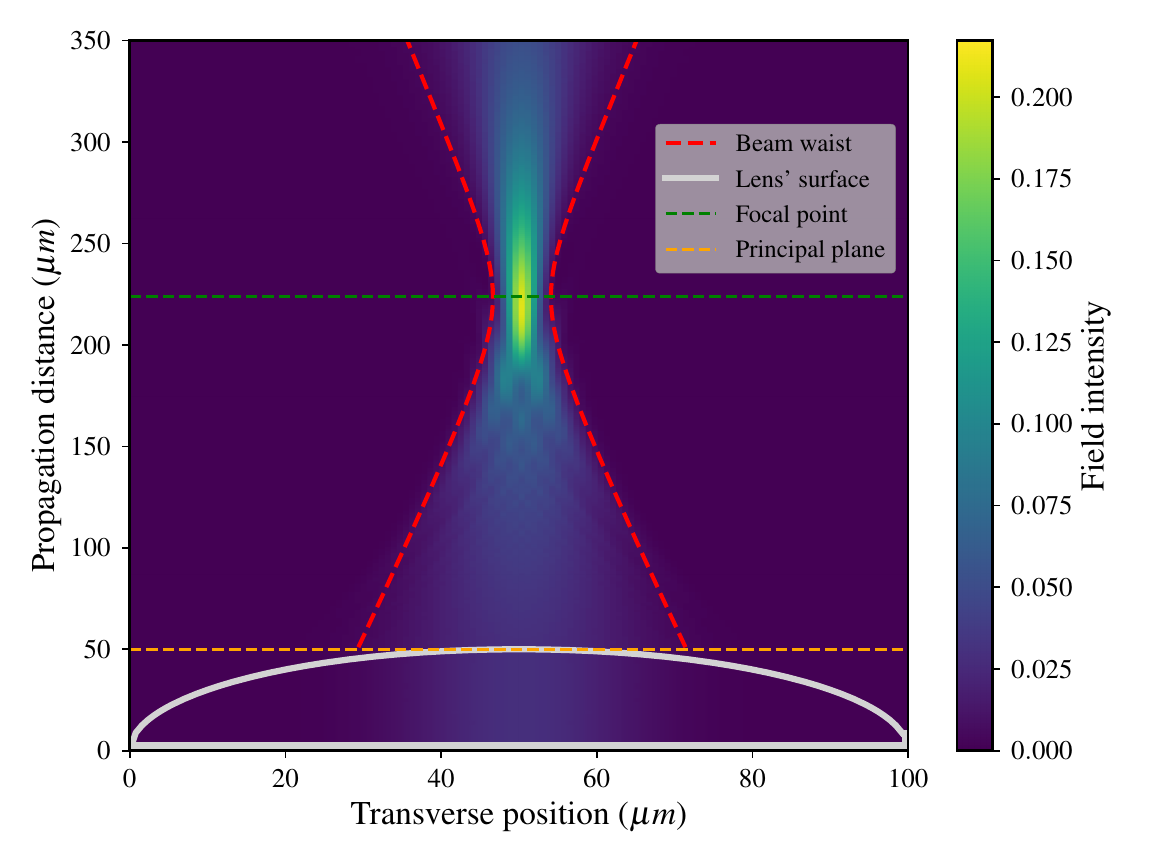}
    \caption{The result of the quantum simulation of a Gaussian beam incident on a plano-convex lens. For this simulation, the maximum phase step was $\Delta_{\mathrm{max}}=0.01$ and the lens was sliced into $10\,000$ layers.}
    \label{fig-simultation-color-plot}
\end{figure}

We ran the simulation with the wavelength, initial Gaussian beam waist, refractive index, radius of curvature, and number of qubits given, respectively, by $\lambda=1\,\mu\mathrm{m}$, $w_0=25\,\mu\mathrm{m}$, $n=1.25$, $R=50\,\mu\mathrm{m}$, and $n_q=7$. The result of this simulation is presented in \cref{fig-simultation-color-plot}.

Summarizing our observations from \cref{fig-simultation-color-plot}, our results reproduce the expected focusing of the beam, and we see the interference of out-of-phase refracted components of the wave before reaching the focus, which is an aberration expected from a thick lens with a spherical surface. Furthermore, it is also known that these aberration effects will decrease if we change the orientation of the lens such that the incident beam hits the convex surface first. To showcase the flexibility of this simulation protocol, we performed the simulation in the reverse orientation and several other cases, such as lenses with parabolic surface curvature, and the results are presented in Appendix~\ref{sec:appendix_simulation}.

Next, we characterized the performance of the quantum simulation. The most important parameter for the performance is the maximum allowed phase step $\Delta_{\mathrm{max}}$ in the block-encodings in \cref{eq-block-encoding-error}, which corresponds to the time step size in the simulation. For benchmarking the effect of $\Delta_{\mathrm{max}}$, we ran simulations keeping all the parameters fixed except for $\Delta_{\mathrm{max}}$, which was varied over the range $[0.001, 0.2]$, and analyzed two metrics: accuracy and probability of success. The expectation is that decreasing $\Delta_{\mathrm{max}}$ leads to higher accuracy and also higher probability of success at the cost of more simulation steps, i.e., larger gate count and simulation time.

\begin{figure}[ht]
    \centering
    \begin{subfigure}{0.45\textwidth}
        \includegraphics[width=\linewidth]{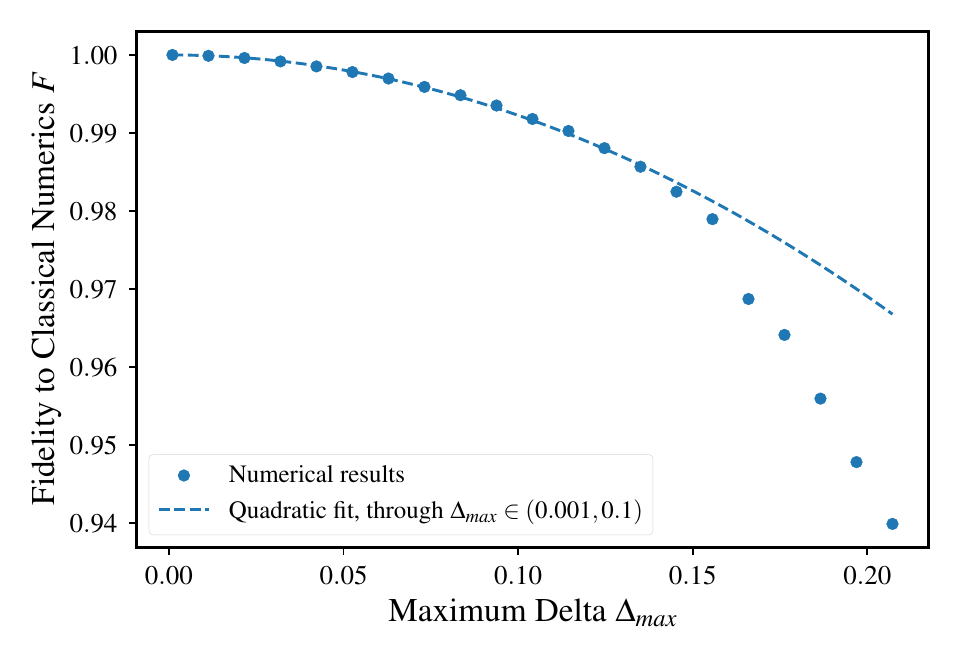}
        \caption{}
        \label{fig-characteristics-fidelity}
    \end{subfigure}
    \hspace{0.089\textwidth}
    \begin{subfigure}{0.45\textwidth}
        \includegraphics[width=\linewidth]{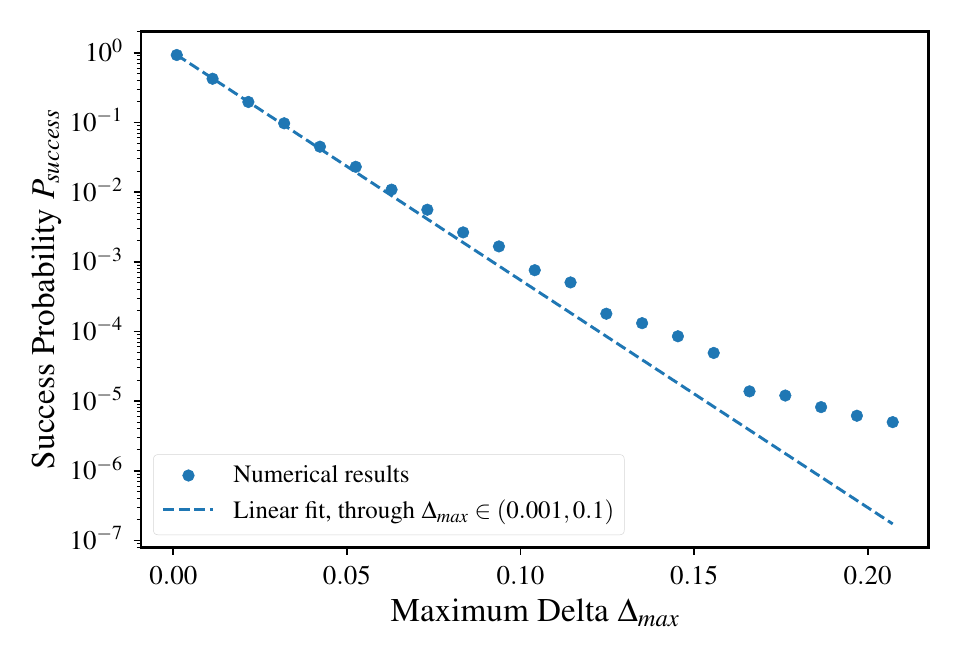}
        \caption{}
        \label{fig-characteristics-success}
    \end{subfigure}
    \caption{Characterization of the error and the performance of the quantum simulation. (a) The fidelity between the resulting wavefunction from the quantum simulation and the classical numerical result obtained by directly multiplying the field with the phases (calculated at $z\approx200\,\mu\mathrm{m}$), post-selecting successful outcomes. The quadratic fit is taken over the range $\Delta_{\mathrm{max}}\in(0.001,0.1)$. (b) The probability of success of the quantum simulation, that is, the probability of all post-selections leading to the desired outcome; the linear fit is taken over the same range. In these simulations, the number of qubits was $n_q=6$.}
    \label{fig-characteristics}
\end{figure}

For measuring the accuracy, we calculated the fidelity between the output state of the quantum simulation and the state resulting from classical calculations at a fixed propagation distance $z\approx200\,\mu\mathrm{m}$. The classical simulation was done by directly multiplying the incident light field by the corresponding phases from the simulation scheme in \cref{eq-time-ordered-evolution-approx} and using classical discrete Fourier transforms for the kinetic energy terms. The reason that the accuracy is defined in this way is to keep other errors such as the midpoint approximation of the time-ordered operator and the split-step error the same between the classical and quantum cases. With that, we isolated the effects of the errors explicitly resulting from the block-encoded phase operators of the quantum simulation protocol. The result is shown in \cref{fig-characteristics-fidelity}.

From analytical calculations (Appendix~\ref{sec:appendix_simulation_error_analysis}), we expect the error to scale as $\epsilon=\mathcal{O}(\Delta_{\mathrm{max}})$. Therefore, the fidelity as a function of $\Delta_{\mathrm{max}}$ behaves as
\begin{equation}
    F=1-\mathcal{O}(\epsilon^2)=1-\mathcal{O}(\Delta_{\mathrm{max}}^2),
\end{equation}
and this is confirmed by fitting a quadratic function to the fidelity in \cref{fig-characteristics-fidelity} for small values of $\Delta_{\mathrm{max}}$. Note that the fidelities are calculated in the case of successful post-selections. To fully characterize the simulation protocol, we should also analyze the success probability. The total probability of success as a function of small $\Delta_{\mathrm{max}}$ is
\begin{equation}
    P_{\mathrm{success}} = 1-\mathcal{O}(\epsilon) = 1-\mathcal{O}(\Delta_{\mathrm{max}}),
\end{equation}
meaning that for small $\Delta_{\mathrm{max}}$ we expect a linear behavior, which is seen in \cref{fig-characteristics-success}. We conclude the discussion by mentioning that, although it is not straightforward to calculate an exact practical value for $\Delta_{\mathrm{max}}$ to achieve high accuracy for a particular simulation problem, it can be said that as long as the probability of success is high, i.e., post-selections are successful, we know that the accuracy will also be high. This makes the success or failure of the post-selection process a measure of the accuracy of the simulation as well. In a real-world application, one can start with a reasonable value of $\Delta_{\mathrm{max}}$ and adapt by tuning it to smaller values if a simulation trial fails.

\section{Conclusion}\label{sec-discussion}

In this work, we discussed how the problem of optical wave propagation in a weakly inhomogeneous medium in the paraxial approximation is reduced to a time-dependent Hamiltonian simulation problem. We then introduced a block-encoding structure that enables the implementation of unitary time evolutions accurate to the first-order approximation in propagation time. The method is highly flexible since the Hamiltonian is programmed in the quantum computer using an amplitude oracle that prepares the wavefunction corresponding to a given phase profile.
The problem of unitary phase operator synthesis is therefore mapped to the state preparation problem.
This modularity allows for the simulation of optical elements with various refractive index distributions, as long as the corresponding phase profiles can be prepared efficiently in a quantum register, and any improvement to the gate complexity of the state preparation part is directly transferred to the block-encoded phase operators.

The simulation algorithm is exponentially efficient in memory consumption, requiring $\mathcal{O}(\log N)$ qubits for a discretized light field with the grid size of $N$ in the transverse plane.
A similar improvement (polylogarithmic scaling with $N$) for the circuit depth is also possible for simulations involving refractive index distributions that are efficiently preparable in quantum registers. This condition is fulfilled by a large class of relevant scientific applications where the refractive index distribution is smooth and bounded in its derivative.
Notably, the number of times that the block-encoding construction is needed for the simulation grows only linearly with the number of time steps.
Therefore, the block-encoded implementation does not add further computational complexity in terms of the simulation time beyond what time discretization already requires for achieving a desired accuracy.
Meanwhile, the block-encoding subroutine at each time step can potentially be implemented using a number of gates polylogarithmic in the grid size $N$.
This efficiency promises quantum advantage for a range of high-precision applications.
We also showcased the algorithm by simulating a simple wave optics experiment, that is, the focusing of a Gaussian beam by a lens.

It must be pointed out that the simulation protocol described in this work will not be efficient if the objective is to extract the complete propagated field for a given propagation distance, because the output is computed as wavefunction amplitudes in a quantum register. Trying to extract the field intensity from this quantum register by direct measurement will inevitably lead to losing the quantum speedup. Reconstructing the output field via state tomography results in the same issue. Instead of direct measurements on the output register, one has to choose a particular observable and only extract information about that. This is, however, not a drawback specific to this protocol but a general consideration for many quantum numerical solvers. Nevertheless, focusing on one observable can still be very beneficial, for example, in an optical design application where the observable can be the coupling efficiency, mode overlap, or a Zernike coefficient that needs to be minimized or maximized for an application.

\begin{figure}[ht]
    \centering
    \includegraphics[width=0.3\linewidth]{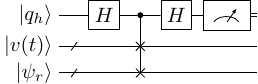}
    \caption{The swap test for estimating the overlap of two quantum states. The qubit $\ket{q_h}$ starts in the $\ket{0}$ state, and the measurement is done in the computational basis.}
    \label{fig-hadamard-test}
\end{figure}

As an example of a concrete readout procedure, we discuss the estimation of the mode overlap of the final output field with a target mode, a common problem appearing in optical applications such as fiber coupling. We propose to estimate the mode overlap using the swap test \cite{dalzell_quantum_2025}.
The swap test takes the register containing the result of the quantum simulator $\ket{v(t)}$ as one of its inputs, while the other input is initialized in a reference state $\ket{\psi_r}$, which for example can be the fundamental mode of a fiber approximated as a Gaussian field. Measuring the test qubit $\ket{q_h}$ after the execution of the swap test subroutine (\cref{fig-hadamard-test}) results in the probabilities $p_0$ and $p_1$ of the qubit being in states $\ket{0}$ or $\ket{1}$, and $p_0-p_1$, which is the expectation value of the Pauli-$Z$ operator on the test qubit, is an estimate of the overlap of the two states $\abs{\braket{\psi_r}{v(t)}}^2$, which is exactly the coupling efficiency to the corresponding fiber mode. Note that since the estimation only needs measuring one qubit, the number of times needed to repeat the process to estimate the overlap to within an error $\epsilon$ is $N_r=\mathcal{O}(1/\epsilon^2)$, independent of the number of qubits in the original register $\ket{v(t)}$. In this way, state tomography is not performed but the relevant information is efficiently extracted. The ability to efficiently simulate the propagation of a light field through optical elements and to estimate design loss functions can greatly enhance iterative design processes and allow for rapid optimization of the design of optical systems.

While the simulation algorithm introduced here, in its current form, is limited to small numerical apertures and cannot yet treat polarization, it shows not only that quantum-enabled optical system design is feasible, but also that exponential speedup is a possibility. We expect our work to motivate the optics community to consider quantum computing as a promising platform for optical simulation and to further investigate the utility of recent developments in quantum algorithms such as block-encoding and quantum signal processing for advances in the field.

\begin{acknowledgments}
This research is funded by the Deutsche Forschungsgemeinschaft (DFG, German Research Foundation) – Project-ID 398816777 – SFB 1375. F.E. and S.D. are affiliated with the Max Planck School of Photonics supported by the German Federal Ministry of Education and Research (BMBF), the Max Planck Society and the Fraunhofer Society.
\end{acknowledgments}

\section*{Author Contributions}
All authors contributed equally to the scientific content of this work. S.D. was a major contributor to developing the simulation codes. All authors read and approved the final manuscript.

\section*{Data and Source Code Availability}
The source code is publicly available in a git repository at \url{https://github.com/BlackWild/qiu}. Snapshots of the repository and the dataset generated and analyzed in this work are permanently stored \cite{davani_2026_19333736_code, davani_2026_19333783_dataset} at \url{https://doi.org/10.5281/zenodo.19333735} and \url{https://doi.org/10.5281/zenodo.19333783}, respectively.

\section*{Competing Interests}
All authors declare that they do not have any financial or non-financial competing interests regarding this work.

\setcounter{figure}{0}
\renewcommand{\thefigure}{\thesection\arabic{figure}} 

\appendix
\counterwithin*{figure}{section} 

\section{Implementation of the conditional phase operator}\label{sec:appendix_implementation}
Here, we provide the details of the implementation of the $U(\theta)$ operator introduced in \cref{sec-results}. The implementation needs $\mathcal{O}(n_q)$ gates, where $n_q$ is the number of qubits of each of the input registers.

The operator is defined in terms of its action on the computational basis states as
\begin{equation}\label{eq:U_definition_implementation_appendix}
    U(\theta) \ket{y}\ket{x} = \begin{cases}
    e^{i\theta} \ket{y}\ket{x}, & \text{for } x=y \\
    \ket{y}\ket{x}, & \text{for } x \ne y
    \end{cases}.
\end{equation}
The goal is to implement this operator by decomposing it into elementary single- and two-qubit gates. The key to its implementation is that the condition $x=y$ in \cref{eq:U_definition_implementation_appendix} is realized when all corresponding qubits of the registers $\ket{x}$ and $\ket{y}$ are in the same state. If we represent the basis states of each register as the tensor product of the underlying qubits, we have
\begin{equation}
    \ket{x} = \ket{x_{n_q-1}}\cdots\ket{x_1}\ket{x_0},
\end{equation}
and similarly
\begin{equation}
    \ket{y} = \ket{y_{n_q-1}}\cdots\ket{y_1}\ket{y_0},
\end{equation}
where $x_j\in \{0,\,1\}$ and $y_j\in \{0,\,1\}$ are the bits of the binary representations of $x$ and $y$. The condition $x=y$ is equivalent to $x_j=y_j$ for all $j$.

The equality of each qubit pair can be checked via elementary unitary operations. To check the condition $x_j=y_j$ we should calculate the flag bit $z_j=y_j \oplus \overline{x_j}$, where $\oplus$ is the XOR logical operator and the overline in $\overline{x_j}$ represents the logical NOT operation. The flag state $\ket{z_j}=\ket{y_j \oplus \overline{x_j}}$ can thus be computed using a CNOT gate acting on the $\ket{y_j}$ qubit with $\ket{x_j}$ as the control qubit and control state $0$,
\begin{equation}
    \mathrm{CNOT}_0 \ket{x_j}\ket{y_j} = \ket{x_j}\ket{y_j \oplus \overline{x_j}} = \ket{x_j}\ket{z_j}.
\end{equation}
Using $n_q$ $\mathrm{CNOT}_0$ gates, we can compute all the flag bits $z_j$; a multi-controlled phase gate $P(\theta)$ with the phase parameter $\theta$ acting on the $\ket{z_j}$ qubits then applies the phase to the state if all the $\ket{z_j}$ qubits are $\ket{1}$, which is equivalent to the condition $x=y$. Finally, the temporarily computed $\ket{z_j}$ states are uncomputed by applying the same CNOT gates again.

\begin{figure}[ht]
    \centering
    \includegraphics[width = 0.5\textwidth]{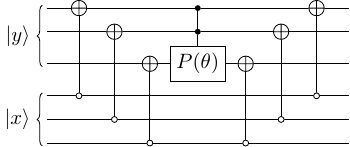}
    \caption{Implementation of the conditional phase operator $U(\theta)$ for $3$-qubit input registers. This is the same circuit as \cref{fig-partial_phase_operator_example}, repeated here for convenience.}
    \label{fig:partial_phase_operator}
\end{figure}

As an example, the implementation for the case of $3$-qubit quantum registers ($n_q=3$) is presented in \cref{fig:partial_phase_operator}. Despite the appearance of the circuit structure, the $U(\theta)$ operator is in fact symmetric with respect to the exchange of its input registers, as is clear from \cref{eq:U_definition_implementation_appendix}.

As we see in this implementation, we need $2n_q$ CNOT gates ($n_q$ to compute the flag bits and $n_q$ to uncompute them) and one multi-controlled phase gate acting on $n_q$ qubits. Because the multi-controlled phase gate can be implemented using $\mathcal{O}(n_q)$ gates \cite{nielsen_quantum_2012}, the total gate complexity of the $U(\theta)$ operator is $\mathcal{O}(n_q)$.

\section{Derivation of the block-encoding}\label{sec:appendix_outcome_calculations}

Here, we discuss the derivations of the results presented in \cref{sec-results} in detail. The discussion includes the block-encoding of the operator
\begin{equation}\label{eq-appendix-block-encoding}
    A(\theta) = I +  2i\,e^{i\frac{\theta}{2}}\sin\frac{\theta}{2} \, \sum_x\abs{\phi(x)}^2 \Pi_x,
\end{equation}
its approximation for $\theta=\Delta\ll 1$
\begin{equation}
    A(\Delta) \approx e^{i\Delta\sum\limits_x\abs{\phi(x)}^2 \ketbra{x}{x}},
\end{equation}
and the case of its iterative application.

Let us start by finding a useful representation of the conditional phase operator in \cref{eq:U_definition_implementation_appendix}. The operator can be rewritten as
\begin{equation}
    U(\theta) = \sum_x \left(\left[ e^{i\theta} \Pi_x + \sum_{y\ne x} \Pi_y \right] \otimes \Pi_x \right),
\end{equation}
where $\Pi_x=\ketbra{x}{x}$ is the projection operator that projects vectors onto the $\ket{x}$ basis state. Using the completeness relation $\sum_y \Pi_y = I$, we have
\begin{align}
    U(\theta)
    &= \sum_x \left[ I + 2i\,e^{i\frac{\theta}{2}}\sin\frac{\theta}{2} \, \Pi_x \right] \otimes \Pi_x\,. \label{eq:u_p_x}
\end{align}
To simplify the notation for the following derivations, we define $U_x = I + 2i\,e^{i\frac{\theta}{2}}\sin\frac{\theta}{2} \, \Pi_x$ and rewrite \cref{eq:u_p_x} as
\begin{equation}\label{eq:U_simplified}
    U(\theta) = \sum_x U_x \otimes \Pi_x.
\end{equation}
This operator does not contain any reference to the $\phi(x)$ state, and therefore it is not capable of applying $x$-dependent phase profiles. We need to involve the $U_\phi$ operator to form the complete block-encoding. $U_\phi$ is defined such that it prepares the $\ket{\phi}$ state from the zero state, $\ket{\phi}=U_\phi\ket{0}$. Note that this definition does not uniquely identify $U_\phi$ because its action on other computational basis states, $U_\phi\ket{j}$, is not specified. However, only the action on the zero state will be relevant in the overall block-encoding construction, and therefore any $U_\phi$ fulfilling $\ket{\phi}=U_\phi\ket{0}$ is sufficient for the implementation. Let us now construct the complete block-encoding unitary $U$ including the $U_\phi$ and $U_\phi^\dagger$ operators acting on the ancillary register
\begin{align}
    U &= \Big( U^\dagger_\phi \otimes I  \Big) \Big( U(\theta) \Big) \Big( U_\phi \otimes I  \Big) \nonumber \\
    &= \sum_x  U^\dagger_\phi U_x U_\phi \otimes \Pi_x.\label{eq-complete-U}
\end{align}
This means that the unitary $U$ is a block-encoding of the following operator
\begin{align}
    A(\theta)
    &= \Big( \bra{0} \otimes I \Big) U \Big( \ket{0} \otimes I \Big) \\
    &= \sum_x \Big( \bra{0} U^\dagger_\phi U_x U_\phi \ket{0} \Big) \Pi_x \\
    &= \sum_x \Bigg( 1  + 2i\,e^{i\frac{\theta}{2}}\sin\frac{\theta}{2} \, \abs{\phi(x)}^2 \Bigg) \Pi_x \\
    &= I +  2i\,e^{i\frac{\theta}{2}}\sin\frac{\theta}{2} \, \sum_x\abs{\phi(x)}^2 \Pi_x,
\end{align}
which is equal to the operator we wanted to construct (\cref{eq-appendix-block-encoding}).
The probability of success of the post-selection, assuming the primary state was initially $\ket{\psi}=\sum_x\psi(x)\ket{x}$, is
\begin{align}
    P_{\mathrm{success}}=\bra{\psi}A^\dagger(\theta) A(\theta) \ket{\psi} &= \sum_x \abs{\psi(x)}^2 \Bigg( 1 - 4 \sin^2\frac{\theta}{2} \, \abs{\phi(x)}^2 \left( 1-\abs{\phi(x)}^2 \right) \Bigg) \label{eq:appendix_prob_of_succ} \\
    &= 1 - 4\Gamma^2\sin^2\frac{\theta}{2} \\
    &\ge 1 - \sin^2\frac{\theta}{2},
\end{align}
where
\begin{equation}
    \Gamma^2 = \sum_x \abs{\psi(x)}^2 \, \abs{\phi(x)}^2 \left( 1-\abs{\phi(x)}^2 \right) \le \frac{1}{4}
\end{equation}
is a constant that depends on the initial state and the applied phase profile.
Furthermore, the wavefunction after post-selection is
\begin{equation}
    \ket{\psi'}=\frac{A(\theta) \ket{\psi}}{\sqrt{P_{\mathrm{success}}}}.
\end{equation}
We can thus say that the operator $U$ in \cref{eq-complete-U} is a $(1,n_q,0)$-block-encoding of $A(\theta)$ in \cref{eq-appendix-block-encoding}, which concludes the first part of our derivation.
Next, we investigate the condition that brings the
$A(\theta)$ operator into the form of a diagonal phase operator.

In the approximated case of $\theta=\Delta\ll 1$, the block-encoded operator is
\begin{equation}\label{eq-block-enc-approx}
    A(\Delta)
    =e^{i\Delta\sum\limits_x\abs{\phi(x)}^2 \ketbra{x}{x}} + \mathcal{O}(\Delta^2),
\end{equation}
therefore, the same operator $U$ is a $(1,n_q,\mathcal{O}(\Delta^2))$-block-encoding of
\begin{equation}
    e^{i\Delta\sum\limits_x\abs{\phi(x)}^2 \ketbra{x}{x}}.
\end{equation}
The error of this implementation, defined using the spectral norm, is
\begin{equation}
    \epsilon = \norm{\frac{A(\Delta)}{\sqrt{P_{\mathrm{success}}}} - e^{i\Delta\sum\limits_x\abs{\phi(x)}^2 \ketbra{x}{x}}} = \mathcal{O}(\Delta^2)
\end{equation}
and the probability of success is
\begin{equation}
    P_{\mathrm{success}} = 1 - \mathcal{O}(\Delta^2) = 1 - \mathcal{O}(\epsilon)
\end{equation}
which concludes the second part of the derivations.

For the final part of the derivations, we investigate the computational complexity of realizing the phase operator
\begin{equation}
    e^{i\sum\limits_x f(x) \ketbra{x}{x}},
\end{equation}
given an arbitrary phase profile $f(x)$, by repeated application of the block-encoding unit in \cref{eq-block-enc-approx}.
In order to use the previous results, the first step is to write $f(x)$ in a form suitable for the block-encoding construction as
\begin{equation}
    f(x) = \alpha\abs{\phi(x)}^2
\end{equation}
for some coefficient $\alpha$ and a normalized wavefunction $\phi(x)$. We assume that $f(x)$ is positive for now, but this requirement will be lifted later. Since $\phi(x)$ is normalized, the coefficient $\alpha$ is calculated as
\begin{equation}
    \alpha = \sum_x f(x).
\end{equation}
The idea is then to write $\alpha=m\Delta$ and repeat the block-encoded operator in \cref{eq-block-enc-approx} $m$ times with the phase parameter $\Delta$.
Since the error at each step is $\mathcal{O}(\Delta^2)$ and the total worst-case error after $m$ iterations is additive \cite{nielsen_quantum_2012}, we have the total error
\begin{equation}
    \epsilon_{\mathrm{tot}} = \mathcal{O}(m\Delta^2) = \mathcal{O}\!\left(\frac{\alpha^2}{m}\right) = \mathcal{O}(\alpha\Delta),
\end{equation}
and the operation that gets applied with the error $\epsilon_{\mathrm{tot}}$ is
\begin{equation}
    \left( e^{i\Delta\sum\limits_x\abs{\phi(x)}^2 \ketbra{x}{x}} \right)^m = e^{i\alpha\sum\limits_x\abs{\phi(x)}^2 \ketbra{x}{x}} = e^{i\sum\limits_x f(x) \ketbra{x}{x}}.
\end{equation}
Furthermore, the total probability of success of $m$ post-selections is
\begin{equation}\label{eq-prob-iterative}
    P_{\mathrm{success}} = \Bigg[1 - \mathcal{O}(\Delta^2) \Bigg]^m = 1-\mathcal{O}(\epsilon_{\mathrm{tot}}).
\end{equation}
Since we have
\begin{equation}
    \abs{\alpha} = \norm{f}_1 = \sum_x \abs{f(x)},
\end{equation}
for a given $f$, the error as a function of the phase and the number of iterations used to realize it is
\begin{equation}
    \epsilon_{\mathrm{tot}} = \mathcal{O}\left(\frac{\norm{f}_1^2}{m}\right).
\end{equation}
In the case of an arbitrary $f$ without the positivity assumption, one can still write down the phase contributions as a sum of positive and negative parts
\begin{equation}
    \sum\limits_x f(x) \ketbra{x}{x} = \sum\limits_{f(x) > 0} \abs{f(x)} \ketbra{x}{x} - \sum\limits_{f(x) < 0} \abs{f(x)} \ketbra{x}{x}
\end{equation}
and implement the two parts using phase steps $\Delta=\norm{f}_1/m$ and $-\Delta$, respectively, which leads to the same total error scaling $\epsilon_{\mathrm{tot}} = \mathcal{O}\left( \norm{f}_1^2/m \right)$.
The total number of gates to implement the iterative case is
\begin{equation}
    \mathcal{O}\Big( m \big(\log N + G \big) \Big),
\end{equation}
where $G$ is the gate complexity of the initializer $U_\phi$ (or $U_\phi^\dagger$).

\section{Error analysis of the simulation}\label{sec:appendix_simulation_error_analysis}

The paraxial-approximated differential equation (\cref{eq-schorodinger-for-propagation}) governing the propagation of an optical wave $v$ (setting $\hbar=c=1$) is
\begin{align}
    i\pdv{}{t}\ket{v(t)}
    &= \hat H(t) \ket{v(t)} \\
    &= \Big[ -\frac{\hat\partial_x^2}{2\tilde{k}} + V(t; \hat x) \Big] \ket{v(t)},
\end{align}
and the solution to this equation starting at $t=0$ and propagating to time $t=T$ is formally written as
\begin{equation}
    \ket{v(T)} = \hat U(T,0) \ket{v(0)} = \mathcal{T} \exp\Big( -i\int_0^T \hat H(t) \, dt \Big) \ket{v(0)},
\end{equation}
where $\mathcal{T}$ denotes time ordering and the potential term $V(t; x)$ encodes the phases that the wave accumulates at each step of the propagation based on the geometry and the refractive index distribution of the space. We assume a one-dimensional transverse axis to match the simulation in \cref{sec-sim}, although most of the derivations are valid in the general case with minor modifications. There are several sources of error at different stages of transforming this formal solution into an expression suitable for a numerical solver, namely
\begin{equation*}
\begin{alignedat}{2}
    \text{exact:}
    &\qquad&
    \mathcal{T}
    &e^{-i\int \hat H(t)\,dt}, \\[1mm]
    \text{time discretization:}
    &\qquad&
    \prod_j
    &e^{-i\hat H(t_{j+1/2})\Delta t}, \\[1mm]
    \text{kinetic-potential split:}
    &\qquad&
    \prod_j
    &e^{-i\hat K\Delta t}
    e^{-i\hat V(t_{j+1/2})\Delta t}, \\[1mm]
    \text{transverse space discretization:}
    &\qquad&
    \prod_j
    &e^{-i\hat K_h\Delta t}
    e^{-i\hat V_h(t_{j+1/2})\Delta t}, \\[1mm]
    \text{quantum circuit implementation:}
    &\qquad&
    \prod_j
    &\hat U_{K_h}
    \hat U_{V_h(j)}.
\end{alignedat}
\end{equation*}
Notably, except for the final implementation stage, the approximations at each stage are common between classical and quantum solvers. In the following, the norm of an operator $\norm{\cdot}$ is defined as the spectral norm.

The first approximation needed is to discretize the time-ordered operator as a sequence of small steps in time by defining the total number of time steps $N_t$, the step size $\Delta t=T/N_t$, and the time grid points $t_j=j\Delta t$. We use a midpoint approximation \cite{blanes_magnus_2009, hochbruck_magnus_2003}, meaning that the Hamiltonian in the time window $[t_j, t_{j+1}]$ is approximated as $\hat H(t_{j+1/2})$. This leads to an approximation of the time evolution operator as
\begin{align}
    \hat U(T,0)
    &= \mathcal{T} \exp\Big( -i\int_0^T \hat H(t) \, dt \Big) \\
    &= \prod_{j=0}^{N_t-1} e^{-i\hat H(t_{j+1/2})\Delta t} + \epsilon_{\Delta t},
\end{align}
where the error is
\begin{align}
    \epsilon_{\Delta t}
    &= \mathcal{O}\Bigg( T\Delta t^2 \Big( \max_t\norm{\partial_t^2 \hat H(t)} + \max_{t}\norm{\comm{\hat H(t)}{\partial_t\hat H(t)}} \Big) \Bigg)\\
    &= \mathcal{O}\Bigg( T\Delta t^2 \Big( \max_t\norm{\partial_t^2 \hat V(t)} + \max_{t}\norm{\comm{\hat K}{\partial_t\hat V(t)}} \Big) \Bigg).
\end{align}
In a simulation, this error quantifies the lens-slicing error and is negligible as long as the material is weakly inhomogeneous.

The next error results from separating the non-commuting kinetic and potential energy parts of the Hamiltonian. Using first-order Trotter splitting \cite{childs_theory_2021}
\begin{align}
    e^{-i(\hat A+\hat B)\Delta t}
    =
    e^{-i\hat A\Delta t}
    e^{-i\hat B\Delta t}
    + \mathcal{O}\Bigg( \Delta t^2 \,\norm{\comm{\hat A}{\hat B}}  \Bigg),
\end{align}
we get
\begin{align}
    \hat U(T,0)
    &= \mathcal{T} \exp\Big( -i\int_0^T \hat H(t) \, dt \Big) \\
    &= \prod_{j=0}^{N_t-1}
    e^{-i\hat K\Delta t}
    e^{-i\hat V(t_{j+1/2})\Delta t}
    + \epsilon_{s} + \epsilon_{\Delta t},
\end{align}
where the accumulated splitting error for the whole propagation is
\begin{equation}
    \epsilon_{s} = \mathcal{O}\Bigg( T\Delta t \,\max_t\norm{\comm{\hat K}{\hat V(t)}} \Bigg).
\end{equation}

We then need to discretize the transverse axis. Discretization affects the field $v$ as well as the kinetic and potential operators. Given a total transverse length $L$, we define the number of discretization points $N_x$, the grid size $\Delta x = L/N_x$, and the sample points $x_j = j \Delta x$. We sometimes use the substitute notation $h=\Delta x$ to keep the equations compact.
After discretization, the action of the second-order partial derivative operator on the sampled field is approximated as
\begin{equation}
    \partial_x^2\ket{v} \approx \frac{v_{j+1} - 2v_{j} + v_{j-1}}{\Delta x^2} + \mathcal{O}(\Delta x ^2),
\end{equation}
which means the norm of the discretized kinetic operator scales as $\norm{K_h} = \mathcal{O}(\Delta x^{-2})$, and the error resulting from the spatial discretization of the kinetic energy propagator is
\begin{equation}
    \norm{e^{-i\hat K_h\Delta t}-e^{-i\hat K\Delta t}} = \mathcal{O}(\Delta t\Delta x^2).
\end{equation}
A reasonable criterion for choosing temporal and spatial discretization grid sizes $\Delta t$ and $\Delta x$ is to balance the global error resulting from the spatial discretization $\mathcal{O}(T\Delta x^2)$ with the leading error term in the Trotter splitting $\epsilon_s= \mathcal{O}(T\Delta t)$, that is to choose
\begin{equation}
    \Delta t = C \Delta x^2 = \mathcal{O}(\Delta x^2),
\end{equation}
for a scalar $C$.
Intuitively speaking, making the spatial grid finer will only result in more favorable total error scaling if the time step size gets smaller accordingly. For this reason, based on the physical parameters of a simulation problem, the grid size $\Delta x$ must not be chosen smaller than needed for capturing the largest spatial frequency components in the simulation, because it will unnecessarily add to the number of time steps without improving the error scaling.
Therefore, in discussing the spatial discretization errors, the parameter $C=\Delta t/\Delta x^2$ is a more practically relevant parameter than $\Delta t$ or $\Delta x$ separately.
Rewriting the kinetic-potential splitting error in terms of the new parameters after discretization, we have
\begin{equation}
    \epsilon_s = \mathcal{O}\bigg( T\Delta t \norm{K_h}\,\max_t\norm{V_h(t)} \bigg) = \mathcal{O}\left( T \left( \frac{\Delta t}{\Delta x^2} \right) \max_t\abs{V(t)} \right) = \mathcal{O}\left( T C \max_t\abs{V(t)} \right).
\end{equation}

Since the kinetic energy propagator $e^{-i\hat K_h\Delta t}$ is now a discretized $N_x\times N_x$ operator which has a simple time-independent diagonal form in the Fourier basis, it can straightforwardly be implemented using a diagonal quadratic phase propagator and QFT subroutines with no extra error, which in total needs $\mathcal{O}(\log^2 N_x)$ quantum gates \cite{cholsuk_efficient_2024}, and the potential energy propagator $e^{-i\hat V_h(t_{j+1/2})\Delta t}$ can be implemented using the block-encoding described in Appendix~\ref{sec:appendix_outcome_calculations}.

The potential propagator at each time step corresponds to the phase $f_h(t;x_j)=-V_h(t;x_j)\Delta t$, and the $1$-norm of this phase is $\norm{f_h(t)}_1 = \norm{V_h(t)}_1 \Delta t$.
In order to analyze the implementation cost, we need to calculate how this norm scales with the discretization parameters. As we will argue in the following, the requirement that $\Delta t$ scales as $\Delta x^2$ makes the norm of $f_h$ small enough for the block-encoding implementation to be efficient.
In the limit of large $N_x$, there is a relation between the average value of a continuous signal and its discretization
\begin{equation}
     \lim_{N_x\to\infty}\frac{1}{N_x} \sum_{j=0}^{N_x-1} \abs{V_h(x_j)} = \frac{1}{L} \int_0^L \abs{V(x)} \, dx,
\end{equation}
thus the norms relate as $\norm{V_h(t)}_1 = \mathcal{O}( \norm{V(t)}_1 / \Delta x )$. Combining this relation with the scaling of the time step with the grid size $\Delta t = C\Delta x^2$ to calculate the norm of the discretized phase function $\norm{f_h(t)}_1 = \norm{V_h(t)}_1 \Delta t$, we have
\begin{equation}
    \norm{f_h(t)}_1 = \mathcal{O}\Big( C \Delta x \, \norm{V(t)}_1 \Big).
\end{equation}
Assuming at each time step we allow for $m$ iterative applications of the block-encoding to realize the phase, the implementation of the potential energy propagator introduces the local error
\begin{equation}
    \epsilon_{b,\mathrm{local}} = \mathcal{O}\Big( \frac{C^2 \Delta x^2 \norm{V(t)}_1^2}{m} \Big) = \mathcal{O}\Big( \frac{C\Delta t \norm{V(t)}_1^2}{m} \Big),
\end{equation}
which leads to the total error
\begin{equation}
    \epsilon_{b} = \mathcal{O}\Big( \frac{TC\max_t\norm{V(t)}_1^2}{m} \Big).
\end{equation}
Comparing this result to the splitting error $\epsilon_s=\mathcal{O}\left( T C \max_t\abs{V(t)} \right)$, one sees that, after discretization, the requirement to keep the global kinetic-potential splitting error small leads to small phases for the potential energy propagator at each time step such that it makes the block-encoding in Appendix~\ref{sec:appendix_outcome_calculations} suitable for the implementation of the propagators. This implementation results in an additive global error that is no worse than the error introduced by discretization after choosing $m=\mathcal{O}(1)$ local iterations. In this case, we have the final result of
\begin{align}
    \hat U(T,0)
    &= \mathcal{T} \exp\Big( -i\int_0^T \hat H(t) \, dt \Big) \\
    &= \prod_{j=0}^{N_t-1}
    \hat U_{K_h}
    \hat U_{V_h(j)}
    + \epsilon_{b} + \epsilon_{s} + \epsilon_{\Delta t},
\end{align}
where $\epsilon_{b} = \mathcal{O}\Big( TC\max_t\norm{V(t)}_1^2 \Big)$ for $m=\mathcal{O}(1)$.

This concludes the analysis of the errors introduced at each stage of the quantum algorithm.
The total numbers of gates for implementing the kinetic and potential energy propagators are, respectively, $\mathcal{O}\big(N_t \log^2 N_x\big)$ and
\begin{equation}
    \mathcal{O}\Big( N_t\log N_x + \sum_j G_j \Big) = \mathcal{O}\Big( N_t \big(\log N_x + \max_j G_j \big) \Big),
\end{equation}
where $G_j$ is the number of gates needed to implement the state initializer $U_{\phi(j)}$ corresponding to the phase profile $V(t_{j+1/2};x)$.
Once all the simulation parameters such as $T$, $L$, $\hat V(t)$, and the number of sample points $N_x$ are fixed, the total error scales as
\begin{equation}
    \epsilon_{\mathrm{tot}}=\epsilon_{b} + \epsilon_{s} + \epsilon_{\Delta t}=\mathcal{O}(\Delta t),
\end{equation}
thus the final fidelity between the expected result and the outcome of the simulator is
\begin{equation}
    F = 1 - \mathcal{O}(\epsilon_{\mathrm{tot}}^2) = 1 - \mathcal{O}(\Delta t^2)
\end{equation}
which is the relation used to analyze the performance of the simulator in \cref{sec-sim}. We conclude this appendix by mentioning that the error analysis done here considers the worst-case scenario using the spectral norm. It is possible that the actual final error is much smaller than the worst case in specific applications.

\section{Simulation of various cases and orientations of lenses}\label{sec:appendix_simulation}

In this section, we present additional lens simulations using the presented quantum algorithm to showcase the flexibility of the block-encoding technique for simulating different optical experiments. In wave optics \cite{saleh_fundamentals_1991}, it is known that the focusing performance of a lens with a finite thickness depends on its shape and geometry, namely the radii of curvature of its surfaces and the orientation of the lens.

\begin{figure}[ht]
    \centering
    \begin{subfigure}{0.49\textwidth}
        \centering
        \includegraphics[width=\linewidth]{figures/wave_propagation-r-nf.pdf}
        \caption{Spherical lens, ``bad'' orientation}
        \label{fig-various-sims-r-nf}
    \end{subfigure}
    \hfill
    \begin{subfigure}{0.49\textwidth}
        \centering
        \includegraphics[width=\linewidth]{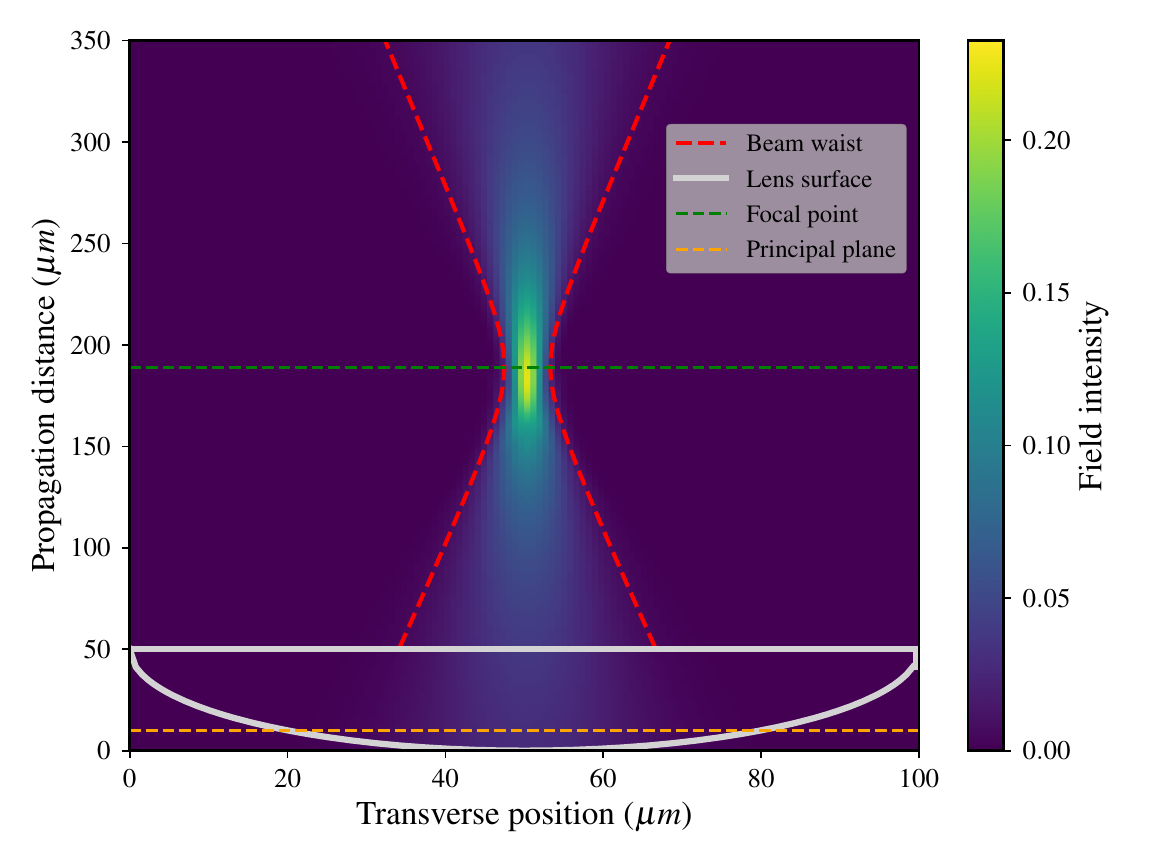}
        \caption{Spherical lens, ``good'' orientation}
        \label{fig-various-sims-nr-nf}
    \end{subfigure}

    \begin{subfigure}{0.49\textwidth}
        \centering
        \includegraphics[width=\linewidth]{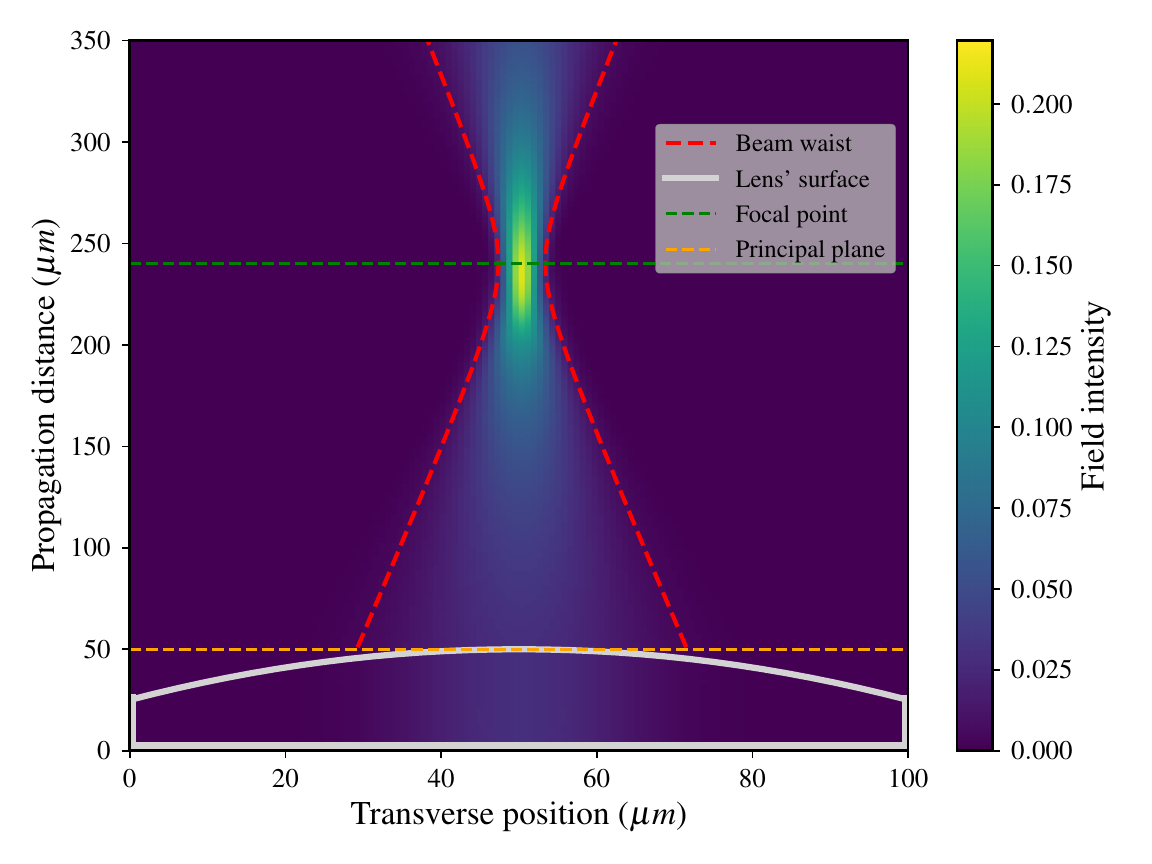}
        \caption{Parabolic lens, ``bad'' orientation}
        \label{fig-various-sims-r-f}
    \end{subfigure}
    \hfill
    \begin{subfigure}{0.49\textwidth}
        \centering
        \includegraphics[width=\linewidth]{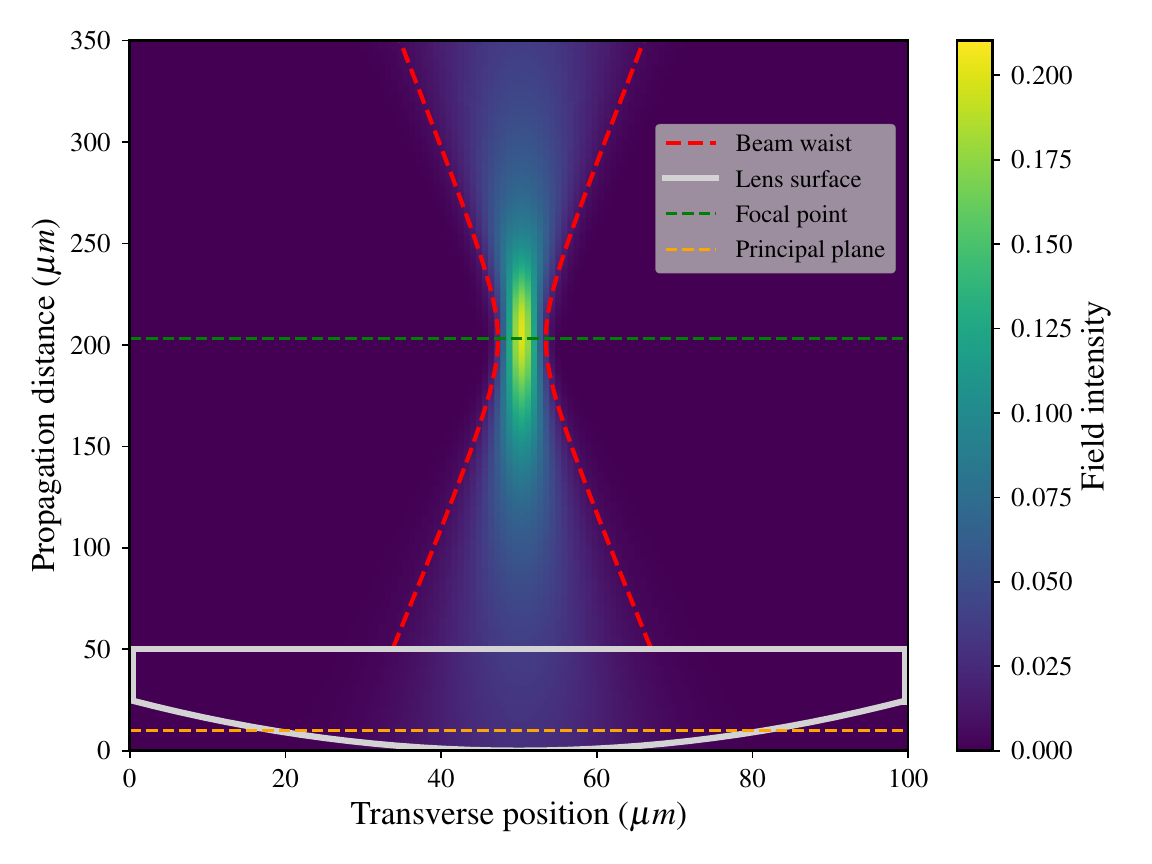}
        \caption{Parabolic lens, ``good'' orientation}
        \label{fig-various-sims-nr-f}
    \end{subfigure}

    \caption{Simulation results of focusing a Gaussian beam incident on various lenses. Panel (a) shows the same simulation as \cref{fig-simultation-color-plot}, repeated here for comparison. The simulation parameters are: the wavelength, initial Gaussian beam waist, refractive index, radius of curvature (on the optical axis), and number of qubits, respectively $\lambda=1\,\mu\mathrm{m}$, $w_0=25\,\mu\mathrm{m}$, $n=1.25$, $R=50\,\mu\mathrm{m}$, and $n_q=7$.}
    \label{fig-various-sims}
\end{figure}

To see these effects, we simulated four different cases. We started with a spherical lens in the orientation where the beam hits the planar surface first (\cref{fig-various-sims-r-nf}) and then simulated the same lens but in the reverse orientation (\cref{fig-various-sims-nr-nf}). Instead of the lens with a spherical surface, we then approximated the curved surface by a parabolic curvature and repeated the simulation in both orientations (\cref{fig-various-sims-r-f,fig-various-sims-nr-f}). The parabolic approximation is chosen such that the curvature on the optical axis is the same for the spherical and parabolic lenses, which means the parabolic lens generally has a smaller curvature at other points of its surface compared to the spherical lens. The general scheme of the simulations is exactly the same as described in \cref{sec-sim}, and the only change is the different phases applied to the incident beam by the lens based on its shape and orientation.

Several effects are expected from and seen in these simulations. It is known \cite{saleh_fundamentals_1991} that the orientation of the lens where the beam hits the curved surface first is considered the ``better'' orientation because the focusing power is distributed over both surfaces, whereas in the other orientation, it is only the curved surface that contributes to the refraction. This leads to two effects. First, the ``bad'' orientation potentially leads to more prominent aberration effects because it amplifies the spherical aberration relatively more. Second, there will be a shift of the focal point because the principal plane (the approximate effective position of the focusing plane when the lens is treated as a thin lens) is located differently depending on the orientation. Comparing the spherical and parabolic lenses in a fixed orientation, we can also see that the focal distance is effectively larger in the parabolic case because of the smaller effective curvature of the parabolic approximation. A parabolic lens also reduces the spherical aberration for the Gaussian beam input.
This shows that the simulation protocol presented in this work offers an intuitive way to run various optical simulations using a quantum computer.

\bibliography{references}

\end{document}